\documentclass[mathleft]{an}
\usepackage{graphicx}
\usepackage{times}
\begin{document}

\Pagespan{789}{}
\Yearpublication{2006}%
\Yearsubmission{2005}%
\Month{11}%
\Volume{999}%
\Issue{88}%

\title{Absorption spectroscopy of gamma-ray burst afterglows:
probing the GRB line of sight}

\author{V. D'Elia \thanks{Corresponding author:
  \email{delia@asdc.asi.it}\newline}
}
\titlerunning{Instructions for authors}
\authorrunning{V. D'Elia}
\institute{
ASI Science Data Centre, Via Galileo Galilei, 00044 Frascati (RM) Italy
\and 
INAF - Osservatorio Astronomico di Roma
}



\abstract{%
  GRB absorption spectroscopy opened up a new window in the study of
  the high redshift Universe, especially with the launch of the {\it Swift}
  satellite and the quick and precise localization of the
  afterglow. Eight-meter class telescopes can be repointed within a
  few hours from the GRB, enabling the acquisition of high
  signal-to-noise ratio and high resolution afterglow spectra. In this
  paper I will give a short review of what we learned through this
  technique, and I will present some of the first results obtained
  with the X-shooter spectrograph.}

\maketitle

\section{Introduction}

Since the discovery of their afterglow emission, it soon became
evident that gamma-ray burst (GRB) spectroscopy could greatly improve
our knowledge of these exciting sources and contribute to unveil the
high redshift Universe. More than ten years later, thanks to the
launch of the {\it Swift} satellite, which disseminates GRB
coordinates in a few tens of seconds, and the implementation of Rapid
Response Mode (RRM) in several facilities, we are able to point
eight-meter class telescopes equipped with high resolution
spectrographs in just a few minutes from the GRB explosion.

Spectroscopy of the GRB line of sight is demonstrating all its
diagnostic power in the following scientific issues: i) to find GRB
redshifts and build the GRB luminosity function; ii) to estimate the
metal content in high redshift galaxies; iii) to characterize the
circumburst environment and to explore the interaction between the GRB
and the surrounding medium; iv) to study the intervening absorbers
along GRB sightlines.

I refer the reader to Fynbo (2011) and Vergani et al. (2009) for
details on items i) and iv), respectively, and I will concentrate on
the other topics, to which the following sections are devoted. The
last section will instead present some of the first results obtained
with the new X-shooter spectrograph.

\section{Metal content in high redshift galaxies}
The study of the metal content of the interstellar medium (ISM) gives
us precious information on the metal enrichment history of the
galaxies. This quantity is in turn linked to the galactic mass
function evolution.

In the past, the study of the metal content in high redshift galaxies
has traditionally relied upon Lyman-break galaxies (LBGs) at $z=3-4$
(see e.g. Steidel et al. 1999) and galaxies that happen to be along
the lines of sight to bright background quasars (or QSOs). However,
both classes present selection effects. LBGs can not be representative
of the true galaxy population, since they are obviously biased toward
high luminosities. Concerning QSOs, their light preferentially probes
the outskirt of the intervening galaxies for a cross section effect.

In this framework, GRBs are a new class of sources whose absorption
spectroscopy provides an independent way to study the metal enrichment
of galaxies at $z>1$.  In addition, using GRBs as torchlights, we are
sure that we are probing the central regions of the galaxies, since
their birthplace are the star-forming regions. Moreover, galaxies are
not selected according to their luminosity, since the probing light
comes from the GRB. Finally, QSOs need at least several hundreds of
million years to form, so in principle GRBs can probe the ISM up to
higher redshifts, as confirmed by the $z=8.2$ GRB\,090423 (Salvaterra
et al. 2009, Tanvir et al. 2009).

Metallicities measured in GRB host galaxies are on average higher than
those found along QSO sightlines (see e.g., Prochaska et
al. 2007). However, the metallicity values are subsolar, varying from
less than $10^{-2}$ to nearly solar values (Savaglio 2010).

\section{The circumburst environment and the need for high resolution
  spectroscopy}
Redshift determination and metallicities of GRBs and their host
galaxies can be derived using ordinary low resolution spectroscopy,
despite a higher resolution is more adequate in order to obtain more
precise column densities and to correctly control the possible
saturation of strong absorption lines.

On the contrary, in order to provide a correct description of the ISM
in general, and the circumburst environment in particular, high
resolution spectroscopy is mandatory. This happens because of several
reasons. 

First of all, the GRB surrounding medium is complex, with many
components contributing to the total absorption; these components can
be separated only using high resolution spectroscopy: this is the only
way to perform an accurate study of the composition, density,
kinematics and physics of the absorbing gas. Fig. 1 (adapted from
Fiore et al. 2005) illustrates this concept by comparing FORS1 low
resolution ($R\sim 1000$) and UVES high resolution ($R \sim 40 000$)
spectra of the same source, GRB\,021004.  It is evident how what
appears as a single absorber in the FORS1 spectrum shows instead a
more complex structure when observed at higher resolutions. This
information allows us also to discriminate between progenitor models,
since in this specific example, the high velocity dispersion and the
constant ionization parameter prefer a Wolf-Rayet wind rather than a
supernova remnant as responsible for the absorption (Fiore et
al. 2005).

\begin{figure}
\includegraphics[width=60mm,height=83mm,angle=-90]{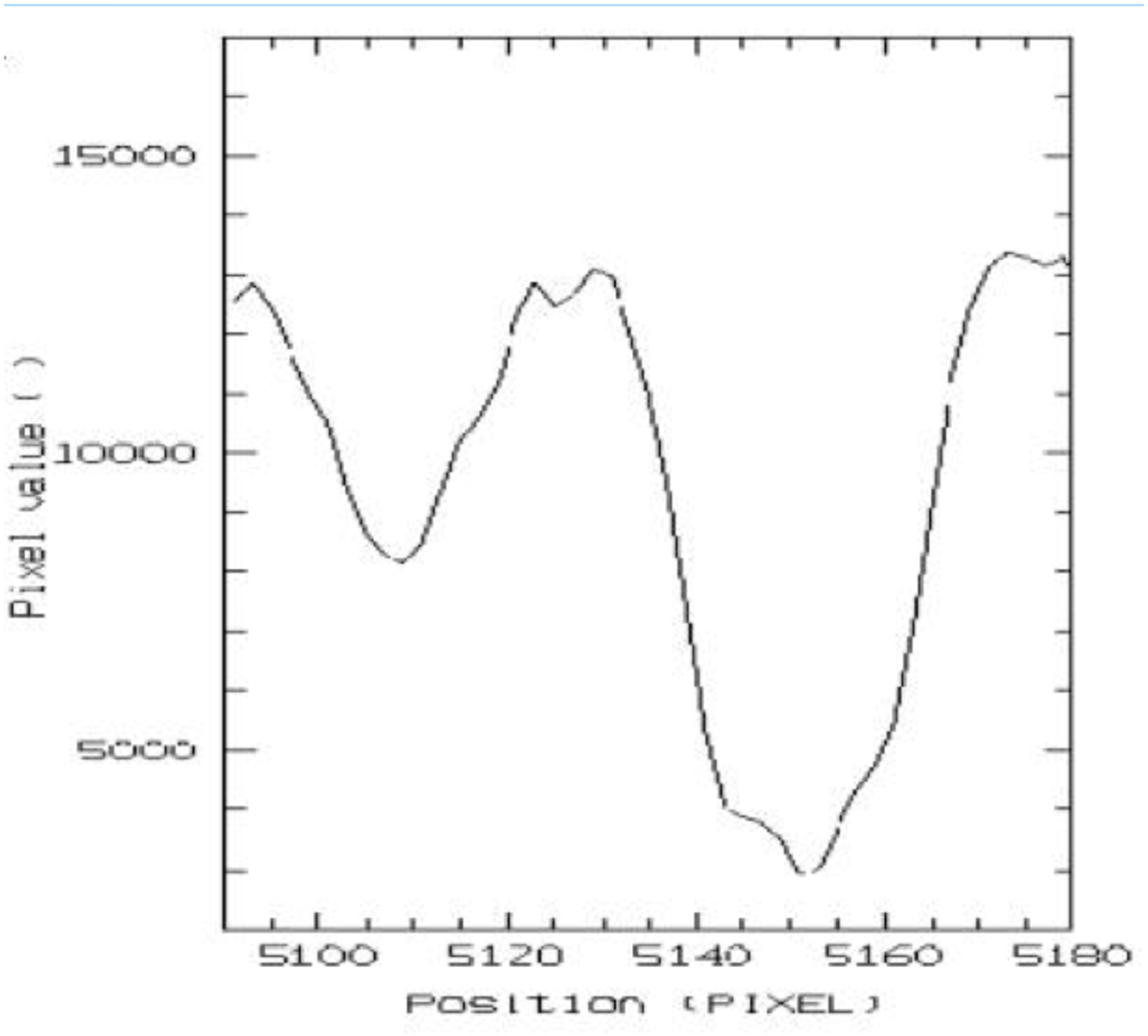}
\includegraphics[width=60mm,height=83mm,angle=-90]{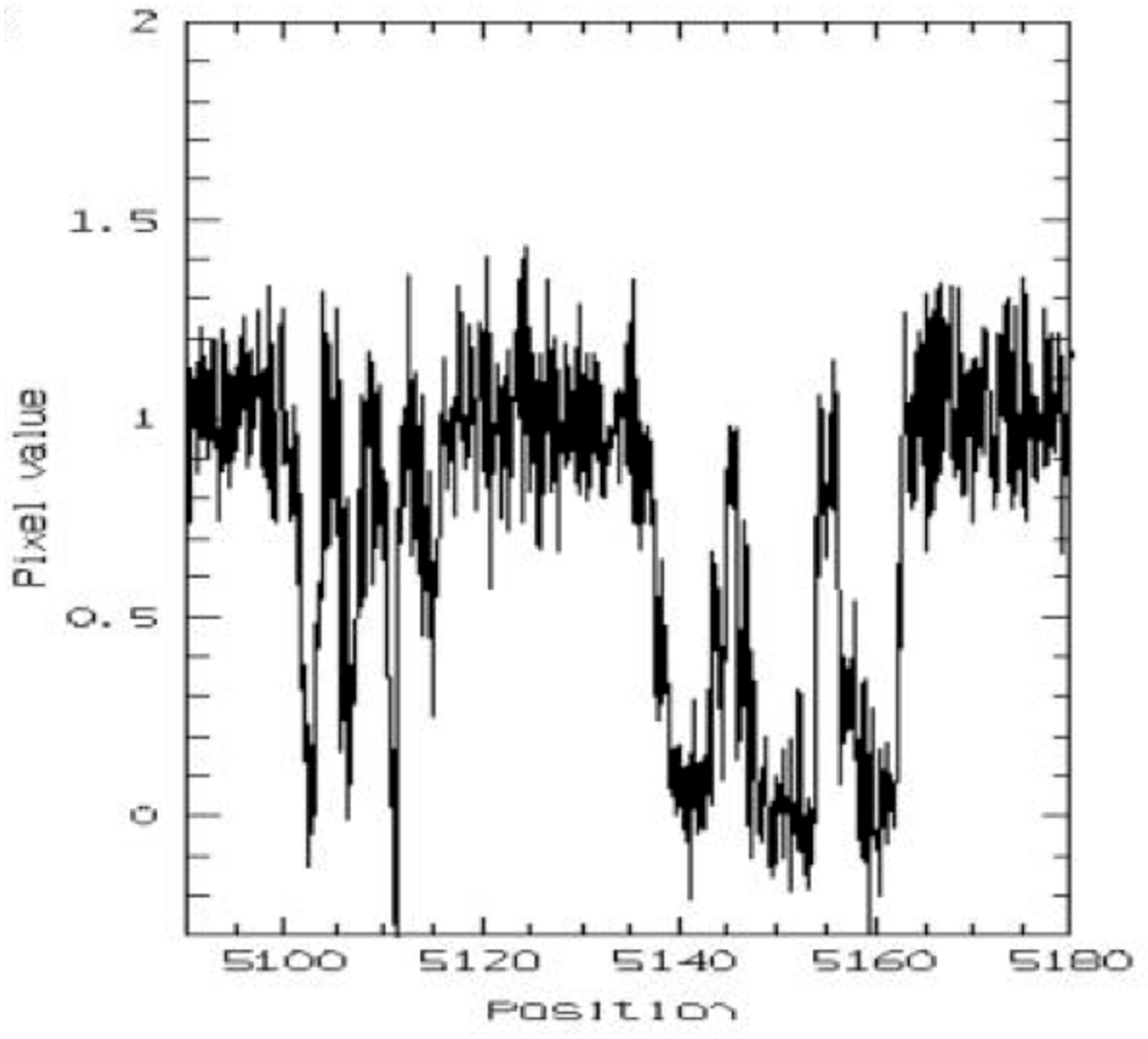}
\caption{The FORS1 low resolution ($R=1000$, top panel) and UVES high
  resolution ($R=40000$, bottom panel) spectra of GRB\,021004, around
  $5140$\AA\, (adapted from Fiore et al. 2005).}
\label{label1}
\end{figure}

Second, some absorbers are close to the explosion site and are
strongly influenced by the GRB output, while others are located far
away. This is evident in absorption components in which high
ionization lines and excited levels are totally absent, reflecting the
scarce influence of the GRB on such absorbers (see e.g. Piranomonte et
al. 2008). The disentangling of these absorbers is only possible using
high resolution spectrographs.

Finally, The level structure of an atom or ion is characterized by a
principal quantum number $n$, which defines the atomic level, and by
the spin-orbit coupling (described by the quantum number $j$), which
splits these levels into fine structure sub-levels. In GRB absorption
spectra, several excited features are detected at the GRB redshift,
due to the population of both $n>1$ and/or $n=1$ fine structure
levels. These transitions are extremely important to gather
information on the interaction between the GRB and the circumburst
environment (see next subsection). To separate these features from the
ground state ones is only possible through high resolution
spectroscopy.

Of course, high resolution is suitable for high luminosity GRBs only,
and a fast reaction to the trigger (now possible thanks to {\it Swift}
and RRM) is needed. In the next subsections I will illustrate two
scientific topics that can be addressed using high resolution, namely,
the excited features and the high ionization lines in GRB absorption
spectra.

\subsection{Excited absorption features}

As mentioned before, fine structure and other excited levels are
commonly observed in GRB afterglow spectra. There is conspicuous
literature on the population of excited states in GRB surrounding
medium and their detection in these spectra (see e.g. Prochaska, Chen
\& Bloom 2006, Vreeswijk et al. 2007 and reference therein). There is
general consensus that these features are produced by indirect UV
pumping by the afterglow, i.e., through the population of higher
levels followed by the depopulation into the states responsible for
the absorption features. This has been proven both by the detection of
variability of fine structure lines in multi-epoch spectroscopy
(Vreeswijk et al. 2007, D'Elia et al. 2009a), and through the column
density ratios of different excited levels when multiple spectra were
not available (Ledoux et al. 2009, D'Elia et al. 2009b). 

Assuming UV pumping as the excitation mechanism, the distance between
the GRB and the absorber can be estimated, because the closer is the
gas to the GRB, the higher are the column densities of the excited
levels. In order to quantify this, a comparison of the observed column
densities with those predicted by a time-dependent, photo-excitation
code is required. This kind of analysis has been performed for 4 GRBs,
and the distance of the absorbers from the burst is in the range
$d=0.3-6$ kpc (Vreeswijk et al. 2007, D'Elia et al. 2009a,b, Ledoux et
al. 2009). This striking result suggests that the power of a GRB
affects a region of gas that is at least a few hundreds pc in size.

\subsection{High ionization lines}

High ionization transitions in GRB spectra are though to be excited
very close to the GRB explosion site. In particular, the column
densities of NV in GRB host galaxies appear to be higher than along
QSO sightlines. This, together with the high ionization potential of
this species, makes the GRB UV flux a privileged candidate for the
production of this ion (Prochaska et al. 2008). If this were the case,
the nitrogen should be located at $\sim 10$ pc from the GRB, a
distance close enough to photo-ionize the NIV and not too high to allow
the photo-production of NVI through the destruction of NV. In this
scenario, a strong variability of these lines is predicted, just like
that of the excited and fine structure levels. On the other hand, Fox
et al. (2008) shows that SIV must be at least at $400$ pc from
GRB\,050730, since its fine structure level is not observed in the
afterglow spectrum. In order to firmly assess if NV is close to the
GRB, a multi-epoch spectroscopy of a $z>2$ GRB is needed to search for
variability of high ionization lines. If this picture will be
confirmed, this will mean that we have access to a region of ISM very
close to the GRB.

\section{X-shooter spectroscopy of GRB\,090926A}

Within this context, X-shooter represents an excellent trade-off
between limiting magnitude accessible for spectroscopy (Mag$_R\sim
21-22$) and achievable spectral resolution ($R \sim 4000 -14000$). In
the framework of the science verification phase program, the afterglow
of GRB090926A was observed with X-shooter $\sim 22$ hr post burst. The
observations consist of 4 different exposures of 600 s each. The
resolution is $R \sim 10000$ and the achieved spectral range is $3000
- 24 800$\AA. The four spectra were co-added after searching for
variability in the absorption features.

In the next subsections I summarize the results obtained analyzing
this spectrum. More details on this work can be found in D'Elia et 
al. (2010).

\subsection{Host galaxy lines and redshift}

The gas residing in the GRB host galaxy is responsible for many of the
features observed in the GRB\,090926A afterglow spectrum. Metallic
features are apparent from neutral (OI, MgI, CaI), low-ionization
(CII, MgII, AlII, AlIII, SiII, SII, CaII, FeII, NiII), and
high-ionization (CIV, NV, OVI, SiIV, SIV) species (see fig. 2). In
addition, strong absorption from the fine structure levels of CII, OI,
SiII, FeII and from the metastable level of NiII is identified,
suggesting that the intense radiation field from the GRB excites such
features. The probed ISM of the host galaxy is resolved into two
components separated by 48 km s$^{-1}$, which contribute to the
absorption system. The wealth of metal-line transitions enables to
precisely determine the redshift of the GRB host galaxy. This yields a
vacuum-heliocentric value of $z = 2.1071 \pm 0.0001$.

\begin{figure}
\includegraphics[width=60mm,height=73mm,angle=-90]{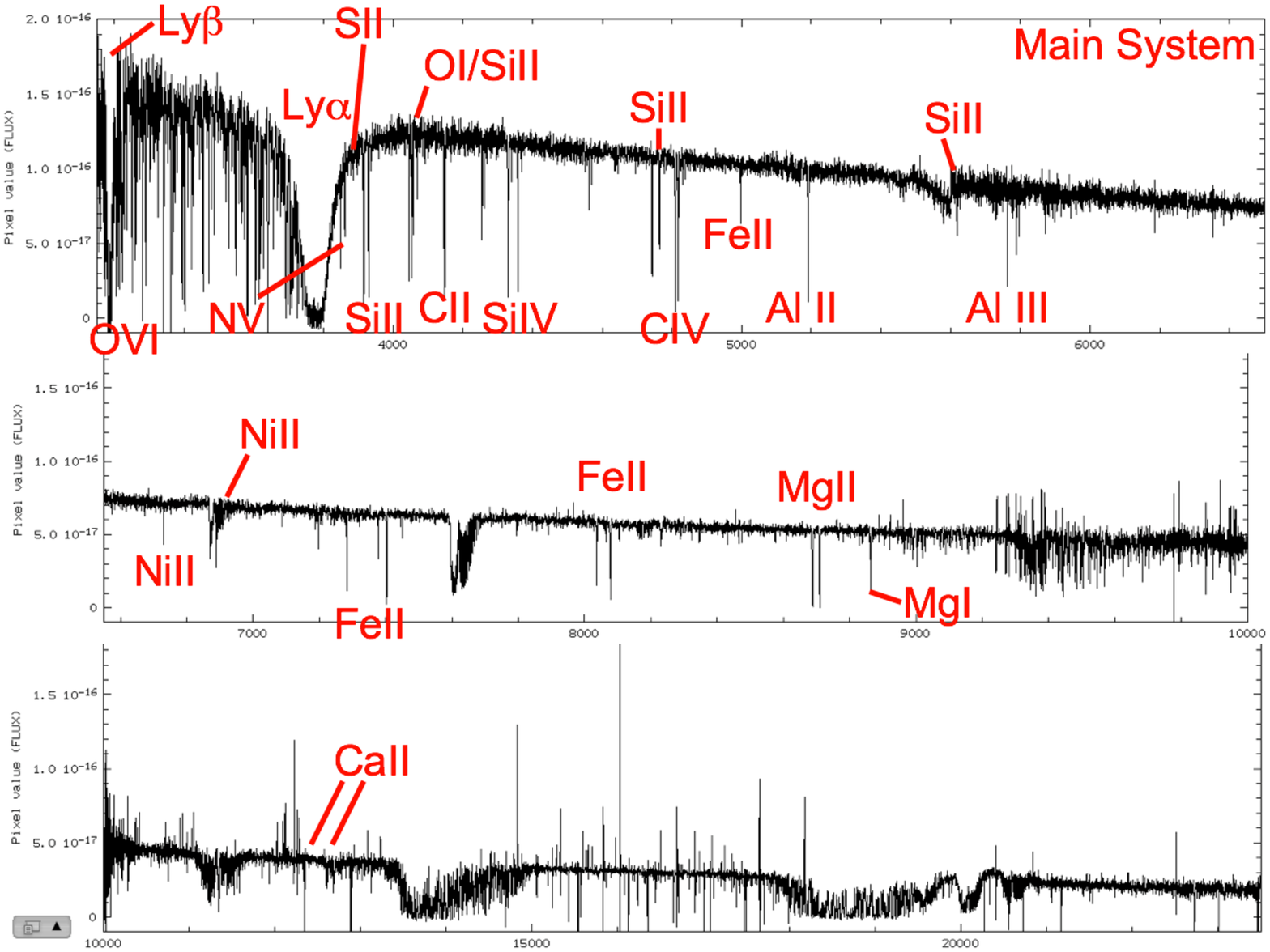}
\caption{The full X-shooter spectrum of GRB\,090926A, together
with all the absorption features identified at the host galaxy
redshift ($z=2.1071$).}
\label{label1}
\end{figure}

\subsection{Metallicities}

The GRB 090926A redshift was high enough to allow the hydrogen
Ly$\alpha$ and Ly$\beta$ lines to enter the X-shooter spectral window.
The hydrogen column density has been constrained using these two
features, and the resulting value is
$log(N_H/$cm$^{-2}$)$=21.60\pm0.07$. Comparing this value with the
metallic column densities, the metallicities of the different species
can be evaluated. Several transitions result to be saturated and were
not taken into account. Despite this, very low metallicity values with
respect to the solar ones, between $4.2 \times 10^{−3}$ and $1.4
\times10^{−2}$, have been derived. These are among the lowest values
ever observed for a GRB host galaxy.

\subsection{Distance between GRB and absorbers}

Comparing the column densities of the ground state and excited lines
of FeII and SiII, the distances between GRB\,090926A and the two
absorbers identified by the detected components can be derived.  For
the redmost component a distance of $d=2.6\pm0.3$ kpc (using FeII) and
$d=2.25\pm0.15$ kpc (using SiII) is computed (see fig. 3). Our best
value is then $d=2.4\pm0.15$ kpc. The bluemost component is far away
from the GRB, at a distance of $\sim 5$ kpc, but in this case the FeII
could not be used because no excited levels were detected, so this
value is computed using SiII only. The GRB\,090926A/absorbers
distances are compatible with what is found for the 4 other GRBs for
which a similar analysis has been performed (See sect. 3.1). This is
further confirmation that the power of a GRB affects a region of gas
that is at least a few hundreds pc in size.

\begin{figure}
\includegraphics[width=60mm,height=60mm,angle=-0]{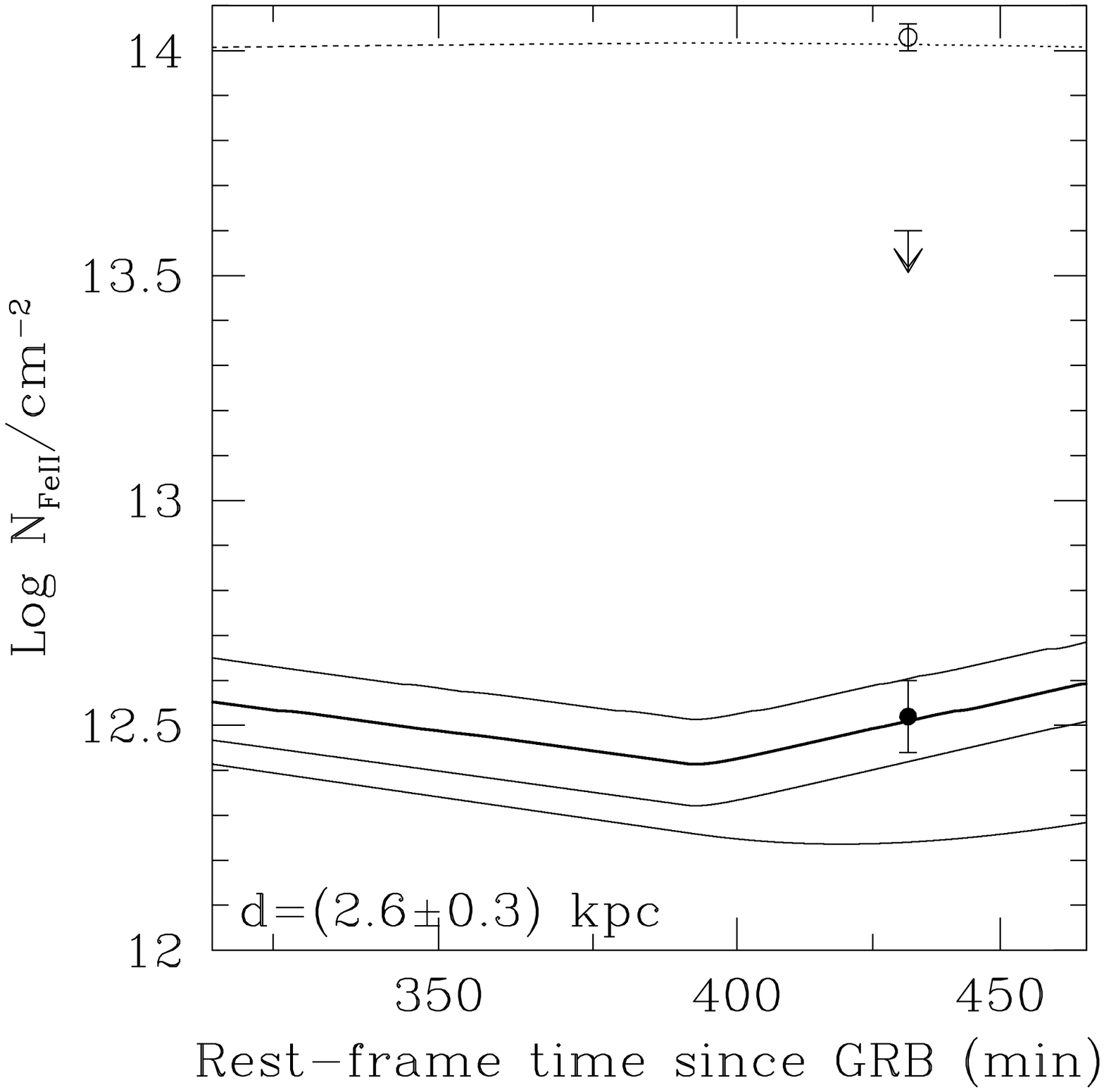}
\includegraphics[width=60mm,height=60mm,angle=-0]{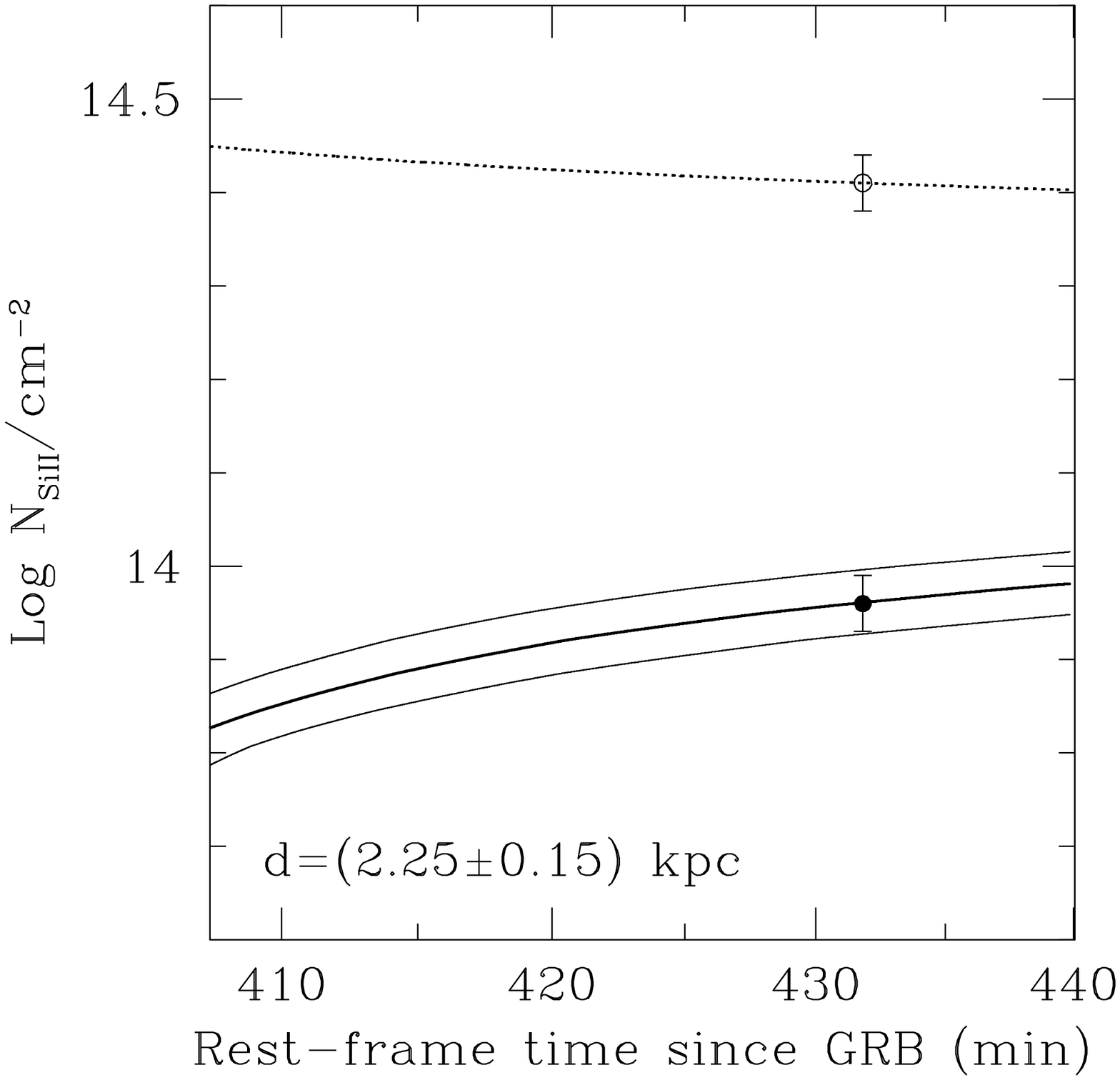}
\caption{Top panel: FeII column densities for the ground level (open
  circle), first fine structure level (filled circle), and first
  excited level (upper limit) transitions for the redmost component of
  GRB090926A. Column density predictions from the time-dependent
  photo-excitation code are also shown. They refer to the ground level
  (dotted line), first fine structure level (thick solid line), and
  first excited level (dashed line) transitions, in the case of an
  absorber placed at $2.6$ kpc from the GRB. The two thin solid lines
  display the models that enclose the fine structure data at $1\sigma$
  level. Bottom panel: same as top panel, but for ground and the first
  fine structure level transitions of SiII.}
\label{label1}
\end{figure}

\subsection{Other features at the host redshift}

No emission lines were detected, but a H$\alpha$ flux in emission of
$9 \times 10^{−18}$ erg s$^{-1}$ cm$^{-2}$ (i.e., a star-formation
rate of $2 M_{\sun}$ yr$^{-1}$), which is typical of many GRB hosts, would have
been detected in our spectra, and thus emission lines are well within
the reach of X-shooter. Similarly, no molecules such as CO and H$_2$
were identified. The upper limit to the H molecular fraction of the
host galaxy ISM is $f < 7\times 10^{−7}$. Again, no diffuse interstellar
bands are present in the X-shooter spectrum.

\subsection{The host galaxy morphology}

A powerful way to infer the nature and the age of objects whose
morphology is unknown is with abundances and abundance ratios (see
Matteucci 2001). This method is based on the fact that galaxies of
different morphological type are characterized by different star
formation histories, and these strongly influence the [X/Fe] versus
[Fe/H] behavior (X being any chemical element). Therefore, if we
compare the measured abundance ratios in the host of GRB\,090926A with
predictions from detailed chemical evolution models, we should be able
to understand the nature of the host. These models show that the host
of GRB\,090926A is probably an irregular galaxy with baryonic mass
$10^8 M_{\sun}$ and evolving with star formation rate (the
inverse of the timescale of star formation) of $0.05$ Gyr$^{-1}$.

\subsection{The extinction curve shape}
The X-shooter's wide spectral coverage enables the search for dust
through the spectral continuum analysis. The flux calibrated spectrum
has been fitted assuming a power-law, with a spectral index of $\beta
= 0.89\pm0.02$. The best fit does essentially not require any
intrinsic extinction because $E_{B−V} < 0.01$ mag adopting a SMC
extinction curve.

\subsection{Intervening systems}

A detailed analysis of the data reveals that at least four intervening
absorbers are present along the line of sight to GRB 090926A. Three of
these systems, those with the highest redshifts ($z = 1.75 - 1.95$),
show absorption from the CIV $\lambda\lambda1548,1550$ doublet, to
which a well-defined HI $\lambda1215$ line corresponds inside the Lyα
forest. The fourth system features instead the MgII$\lambda\lambda
2796,2803$ doublet and (marginally) the MgI$\lambda2852$ line, at the
redshift of $z=1.25$.

\section{Conclusions}

GRB absorption spectroscopy is a key tool to study both the GRB
physics and the high redshift Universe. X-shooter is at the forefront
in providing excellent datasets for these kind of science.




\end{document}